\documentclass{cimento}

\newcommand{\Cf}{$^{252}$Cf}
\newcommand{\Qy}{$Q_{y}$}
\newcommand{\Er}{$E_{r}$}
\newcommand{\BaF}{BaF$_2$}
\newcommand{\Ne}{$N_{e^-}$}

\usepackage{graphicx}  

\title{Extending direct measurements of argon nuclear recoils into the sub-keV regime with ReD and ReD+.}

\author{N.~Pino\thanks{pino@lns.infn.it.}}
\instlist{\inst{} INFN, Laboratori Nazionali del Sud  - Catania, Italy}

\begin{document}

\maketitle

\begin{abstract}
Direct searches for dark matter in the form of WIMPs with argon-based detectors require precise measurements of the ionization yield \Qy\ for nuclear recoils at low energies.  
Prior to this work, direct experimental data were available only above 6.7 keV, leaving a critical gap in the energy region most relevant for low-mass WIMP searches.  
The Recoil Directionality (ReD) experiment addressed this limitation by measuring the argon \Qy\ for nuclear recoils between 2 and 10 keV using a dual-phase TPC irradiated with neutrons from a \Cf\ fission source.  
The results extend existing direct measurements to lower energies, show consistency with previous data above 7 keV, and indicate an enhanced ionization yield at low recoil energies.  
These measurements provide essential input for next-generation argon-based dark matter searches and directly motivate the upgraded ReD+ phase, designed to further extend sensitivity into the sub-keV recoil-energy regime.

\end{abstract}

\section{Introduction}
Dark matter in the form of weakly interacting massive particles (WIMPs) is actively searched for using a variety of detection technologies~\cite{Bertone2005,darkmatter}. Among these, dual-phase time projection chambers (TPCs) filled with noble elements such as xenon or argon are widely employed due to their scalability, intrinsic radiopurity, and powerful background discrimination capabilities~\cite{Aprile2018,DS_50_532}.

In these detectors, a WIMP may interact via elastic scattering off a target nucleus. The deposited energy is measured through prompt scintillation in the liquid phase and a delayed electroluminescence signal in the gas phase, produced by ionization electrons drifted by an applied electric field into the gas volume above the liquid.

Argon has been selected as the target medium for a series of increasingly large TPC-based experiments developed by the Global Argon Dark Matter Collaboration. Its current flagship experiment, DarkSide-20k, is under construction at INFN Laboratori Nazionali del Gran Sasso, Italy~\cite{Aalseth2021}.

Compared to xenon, argon offers advantages for detecting nuclear recoils (NRs) induced by WIMPs with masses of a few GeV, a mass range that has received growing attention in recent years~\cite{Chepel2013,DS_50_lm_searches}. While $\mathcal{O}(100\ \mathrm{GeV})$ WIMPs typically produce recoil energies of several tens of keV in argon, lighter WIMPs with masses of order $\mathcal{O}(1\,\mathrm{GeV})$ are expected to yield nuclear recoils of only a few keV, posing a significant experimental challenge~\cite{Ref_DS20k_Nature}. At such low energies, scintillation in argon becomes too weak to be efficiently detected, making ionization the primary observable. However, the ionization yield in argon for NRs below approximately 7~keV remains poorly constrained. The DarkSide-50 Collaboration addressed this limitation by developing a response model based on the Thomas–Imel box model~\cite{Thomas:1987ek}, which describes the ionization yield \Qy\ — defined as the number of ionization electrons per unit deposited energy — down to about 0.5~keV~\cite{Ref_Masato,PhysRevD.104.082005}. The model parameters were constrained through a global fit combining Monte Carlo simulations, DarkSide-50 neutron calibration data (AmBe and AmC), and direct measurements reported by Joshi et al.~\cite{Joshi:2014fna} and by the ARIS~\cite{Agnes:2018mvl} and SCENE~\cite{Cao:2015ks} experiments, extending down to 7~keV.

The absence of direct measurements at lower energies leaves the choice of nuclear stopping-power description uncertain. Models such as those by Ziegler~\cite{Ziegler}, Molière~\cite{Moliere}, and Lenz–Jensen~\cite{Lenz,Jensen} reproduce the data used in the model construction but yield substantially different \Qy\ predictions below 5~keV.

Direct measurements in the sub-5~keV region are therefore essential to discriminate among stopping-power models and to constrain the detector response where their predictions diverge most strongly.

Within the DarkSide-20k Collaboration, the Recoil Directionality (ReD) experiment was conceived and operated at INFN Sezione di Catania, Italy, to further constrain \Qy\ through direct measurements at low recoil energies~\cite{Agnes:2025rxi}. ReD measured the argon ionization yield in the 2--10~keV recoil-energy range, where Thomas–Imel model predictions are particularly sensitive to the choice of nuclear screening function~\cite{PhysRevD.104.082005}. It employed a two-body kinematic technique to select nuclear recoils in the region of interest, as described in Sects.~\ref{sec:concept} and~\ref{sec:analysis}.

Building on the concepts and results established by ReD, the upgraded phase, ReD+, is designed to improve the experimental configuration and extend these measurements to even lower recoil energies, further strengthening the ReD program. An overview of the ReD+ phase is given in Sect.~\ref{sec:redplus}.

\section{Conceptual design and experimental setup}\label{sec:concept}

The ReD experimental setup was designed to produce argon nuclear recoils (NRs) in the 2–10~keV energy range, as required to achieve the goals of the project. 
This is accomplished by inducing elastic neutron scattering on argon nuclei and determining the recoil energy from the scattering kinematics.

Neutrons are provided by a \Cf\ source. Its spontaneous fission (SF) spectrum extends up to approximately 13~MeV, with a mean energy of 2.3~MeV, making it well suited to populate the recoil-energy region of interest. 

The \Cf\ source, with an activity of approximately 1~MBq, is housed inside boron-loaded polyethylene shielding together with two \BaF\ scintillator crystals used as fission-event taggers. 
These detectors record the prompt $\gamma$ rays emitted in SF, providing the start time for the neutron time-of-flight (ToF) measurement. 
The shielding also includes a polyethylene conical collimator with a $2.6^{\circ}$ opening angle to define the neutron beam. 
The ReD dual-phase liquid argon TPC (LAr TPC) is positioned along the collimated beam at a distance of about 90~cm from the source.

\begin{figure}[tbp]
  \centering
  \includegraphics[width=0.9\textwidth]{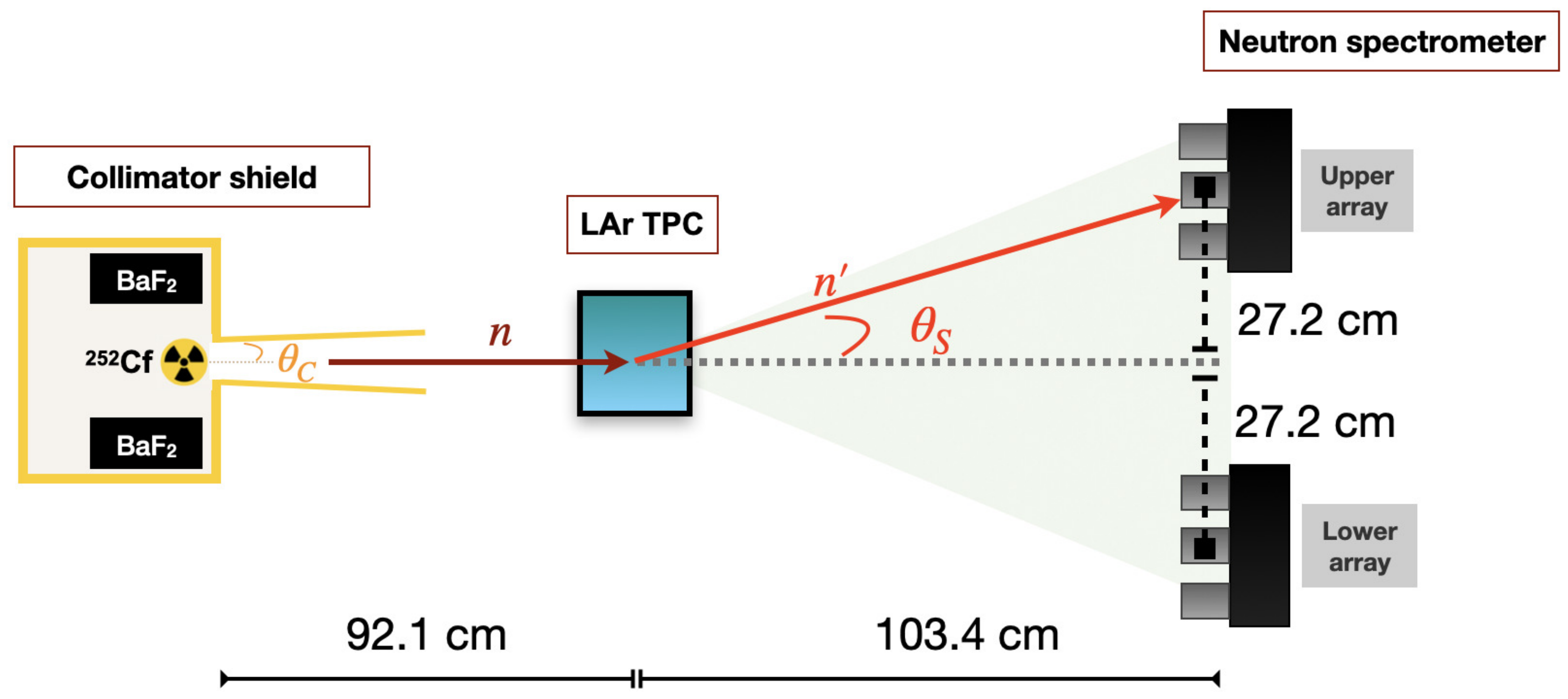}
  \caption{Sketch (not in scale) of the ReD experimental apparatus. From left to right, the setup includes the \Cf\ neutron source surrounded by \BaF\ detectors inside the collimator shield, the dual-phase TPC, and the neutron spectrometer. The arrows indicate a neutron emitted from the source which undergoes (n,n') interaction with Ar in the TPC and is eventually scattered within the acceptance of the neutron spectrometer. See text and Ref.~\cite{Agnes:2025rxi} for more details.}
  \label{fig:setup}
\end{figure}

After scattering in the TPC, neutrons are detected by a downstream spectrometer located approximately 100~cm from the TPC, as schematically shown in Fig.~\ref{fig:setup}. The spectrometer consists of 18 1-inch EJ-256 plastic scintillators (PScis), manufactured by Scionix and selected for their pulse-shape discrimination (PSD) capability for neutron/$\gamma$ separation. 
They are arranged in two $3\times3$ matrices to preserve up-down symmetry and are placed at fixed positions corresponding to scattering angles $\theta_{S}$ between $12^{\circ}$ and $17^{\circ}$. 
This configuration allows the selection of nuclear recoils in specific energy intervals according to scattering kinematics. 
The 1-inch detector diameter provides sufficient spatial granularity for an accurate determination of the interaction position in the spectrometer and, consequently, improved precision on the scattering angle.

In addition to constraining $\theta_{S}$, the PScis provide the stop signal for the ToF measurement of scattered neutrons. 
Combining the start signal from the \BaF\ taggers with the stop signal from the spectrometer enables a precise determination of the neutron ToF, from which the scattered neutron kinetic energy $K_n$ is reconstructed. 
Over the $\sim$200~cm flight path, the neutron ToF ranges from 40 to 160~ns. 
A ToF resolution of about 1~ns (rms) was achieved, corresponding to a $K_n$ resolution of approximately 5\%~\cite{Agnes:2025rxi}.

The argon nuclear recoil energy $E_r$ is then determined event by event from the measured $\theta_{S}$ and neutron kinetic energy $K_n$ as

\begin{equation}
E_{r} = 2 K_{n}\frac{m_n m_{Ar}}{(m_n + m_{Ar})^2} (1 - \cos\theta_{S}),
\label{eq:recoil_energy}
\end{equation}

where $m_n$ and $m_{Ar}$ denote the neutron and argon masses, respectively. 
Within the selected angular range ($12^\circ$–$17^\circ$), the \Cf\ neutron spectrum produces argon NRs between approximately 2 and 10~keV in the TPC.

The ReD LAr TPC features a compact cubic active volume of $5\times5\times6\,\mathrm{cm}^3$, bounded at the top and bottom by transparent ITO-coated quartz electrodes. 
A uniform electric field of approximately 200~V/cm drifts ionization electrons toward a 7-mm-thick gas layer above the liquid, where they are accelerated to produce the delayed electroluminescence signal.

Both the prompt scintillation (S1) and delayed electroluminescence (S2) signals are detected by two cryogenic silicon photomultiplier (SiPM) tiles~\cite{cryoSIPM} located above and below the active volume, behind the transparent electrodes. Each tile hosts 24 custom-made SiPMs arranged in a $4\times6$ array.

The SiPMs in the top tile are read out individually to precisely reconstruct the position of the S2 signal, maximizing the $x$--$y$ spatial resolution of the event. In contrast, those in the bottom tile are grouped in sets of six and read out as four summed channels, since fine granularity is not required in that plane. The $z$ coordinate is determined from the electron drift time, measured as the time difference between the prompt S1 signal and the delayed S2 signal produced in the gas phase.

Since argon scintillation and electroluminescence light are emitted in the vacuum ultraviolet at 128~nm, all internal surfaces are coated with tetraphenyl butadiene (TPB) to wavelength-shift the photons to approximately 420~nm, matching the spectral sensitivity of the photosensors.

Further details on the TPC and the full experimental setup can be found in Refs.~\cite{Agnes:2025rxi,Ref_Naples}.

\section{Data analysis and results}\label{sec:analysis}

The data set used in this analysis was collected over a three-month period, from January to March 2023.

Data acquisition relies on a coincidence trigger formed by signals from at least one \BaF\ detector and one PSci in the spectrometer within a 256~ns time window. Owing to the small scintillation signals expected from low-energy nuclear recoils (S1 $\leq$ 20 photoelectrons), the TPC is operated in follower mode and is therefore excluded from the trigger logic.

The event selection procedure and analysis framework follow the approach described in Ref.~\cite{Agnes:2025rxi}. Candidate neutron-induced events are identified by combining ToF measurements with pulse-shape discrimination in the PScis, which efficiently suppresses $\gamma$-ray backgrounds. The corresponding TPC waveforms are then analyzed to identify associated S1 and S2 signals. The pulse-finding algorithm is fully efficient for S2 signals corresponding to approximately 4~$e^-$.

Events are selected by requiring a single well-defined S2 pulse, possibly accompanied by a weak S1, occurring within a time window compatible with the electron drift time and reconstructed within the central $4\times4$~cm$^2$ region of the TPC. For selected events, the S2 signal measured in photoelectrons (PE) is converted to the corresponding number of extracted electrons \Ne\ using an ionization gain of $g_2 = (18.56 \pm 0.71)\,\mathrm{PE}/e^-$, as determined from calibration data.

The accuracy of the reconstructed \Er\ and \Ne\ is validated using Monte Carlo simulations processed through the full analysis chain~\cite{Ref_LIDINE24}. Both observables are found to be essentially unbiased, with relative resolutions improving at higher energies and charges~\cite{Agnes:2025rxi}.

In total, approximately 800 nuclear-recoil candidate events are identified in the 1–10~keV energy range. Monte Carlo simulations indicate that about 50\% of this sample originates from single elastic (n,n') scattering within the TPC, while the remainder is predominantly due to neutron multiple scatterings. These multi-scatter events are treated as background, as they do not preserve the kinematic relationship between recoil energy and scattering angle given by Eq.~\ref{eq:recoil_energy}. Roughly 70\% of the candidate sample consists of S2-only events, as expected since the S1 signal associated with genuine low-energy NRs is typically too small to be efficiently identified.

\begin{figure}[tbp]
  \centering
  \includegraphics[width=0.9\textwidth]{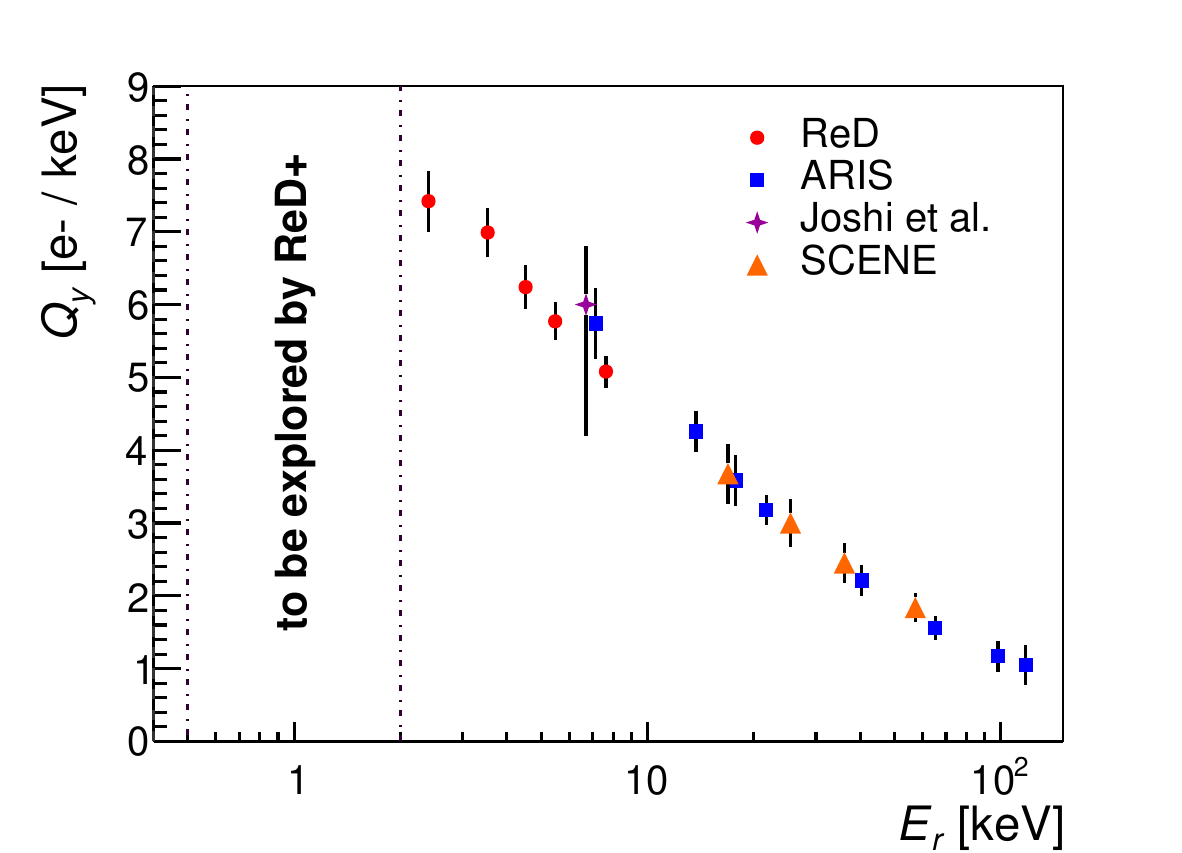}
  \caption{Ionization yield measured in this work for nuclear recoils in the 2–10 keV energy range, including combined statistical and systematic uncertainties. For comparison, literature data up to 120 keV from Joshi et al.~\cite{Joshi:2014fna}, ARIS~\cite{Agnes:2018mvl}, and SCENE~\cite{Cao:2015ks} are also shown. The 6.7 keV data point from Joshi et al. has been rescaled following the procedure described in Ref.~\cite{PhysRevD.104.082005}. The energy region below 2 keV will be explored by the forthcoming ReD+ project (see Sect.~\ref{sec:redplus}).}
  \label{fig:results}
\end{figure}

The ionization yield \Qy\ is extracted in five nuclear-recoil energy bins using unbinned likelihood fits to the observed \Ne\ distributions. The signal component is modeled with a Gaussian function, while a flat term accounts for the contribution from multiple-scattering background events. As shown in Fig.~\ref{fig:results}, the resulting \Qy\ measurements span recoil energies between 2 and 10~keV, extending direct experimental coverage below the previously reported limit of 6.7~keV. Above 7~keV, the results are consistent with existing measurements, whereas at lower energies the ionization yield is observed to increase relative to commonly used extrapolations.

The ReD data were incorporated into a global fit following the approach of Ref.~\cite{PhysRevD.104.082005}, together with DarkSide-50~\cite{DS_50_532}, ARIS~\cite{Agnes:2018mvl}, and SCENE~\cite{Cao:2015ks} results within the Thomas–Imel box model~\cite{Thomas:1987ek}. 
The combined dataset favors the nuclear stopping power description of Lenz and Jensen~\cite{Lenz,Jensen} relative to the Ziegler model~\cite{Ziegler}, the latter predicting an approximately constant trend below 5 keV~\cite{PhysRevD.104.082005}. These results provide input to global response-model fits and help guide the optimization of future argon-based detectors~\cite{newpaper}.


\section{The ReD+ phase}\label{sec:redplus}
Encouraged by the successful results of ReD, the upgraded ReD+ configuration has been conceived to extend the experimental reach to lower NR energies while preserving the detector concept and optimizing its key design features, thereby pushing the sensitivity well below the current few-keV threshold~\cite{Agnes:2025rxi,Ref_LIDINE24}.

The upgraded detector will employ a new dual-phase liquid argon TPC with an enlarged active volume, designed to minimize passive materials and consequently reduce backgrounds associated with neutron multiple scattering. The nearly finalized technical design includes a longer drift length for improved signal separation and a higher $g_2$ to enhance sensitivity to small S2 signals.

To further improve sensitivity to low-energy recoils, the neutron spectrometer will be operated at increased source-to-detector distances and at smaller scattering angles, $\theta_S \approx 6.5^\circ$–$10^\circ$. At these reduced angles, the kinematics correspond to lower nuclear recoil energies, enabling reconstruction in the sub-keV range (approximately 0.6–1.1 keV).

The neutron interaction rate will be increased by replacing the current \Cf\ source with a higher-activity source of about 3~MBq, and by adding two additional $3\times3$ PScis arrays to the spectrometer in a left--right configuration. This extension complements the existing up--down geometry and increases the total number of scintillators to 36, effectively doubling the solid-angle coverage.

Additional mitigation of multiple-scattering backgrounds from neutrons bypassing the collimator or leaking from the front side will be achieved through thicker, improved shielding. A pilot data-taking campaign incorporating these upgrades, while still using the original ReD TPC, is currently taking place at INFN Laboratori Nazionali del Sud in Catania. This will be followed by a full experimental campaign with the optimized TPC, targeting nuclear-recoil energies down to approximately 0.5~keV as shown in Fig.~\ref{fig:results}.

At a later stage, the \Cf\ source will be replaced by a Deuterium–Deuterium neutron generator from the University of São Paulo. This generator produces quasi-monoenergetic 2.4~MeV neutrons via the d(d,$^3$He)n reaction, with a neutron flux exceeding $10^6$~n/s. The associated $^3$He nucleus will be detected by a dedicated silicon detector, enabling event-by-event neutron tagging. This configuration is expected to enable precise measurements of the ionization yield \Qy\ down to \Er\ of approximately 0.2~keV while substantially suppressing background contributions.

\section{Conclusions}
The ReD experiment has directly measured the ionization yield \Qy\ in liquid argon for nuclear recoils in the few-keV energy range, independently of model assumptions. 
These results address the long-standing lack of experimental data relevant for the calibration of large-scale dark matter detectors. 
The observed increase in \Qy\ at lower recoil energies has important implications for sensitivity estimates in upcoming searches for low-mass WIMPs, as well as for studies of coherent elastic neutrino–nucleus scattering.

The planned ReD+ phase builds on these results and is designed to extend the experimental reach into the sub-keV regime. This will provide key input for optimizing next-generation argon-based detectors and improving theoretical descriptions of ionization and recombination processes in liquid argon.


\acknowledgments
The author gratefully acknowledges the support of the INFN Laboratori Nazionali del Sud and of the INFN Catania Section, as well as the contributions of all members of the ReD group.

This work has been supported by the PRIN2022 grant 2022JCYC9E, call for tender No. 104 published on 
2.2.2022 of the Italian Ministry of University and Research (MUR) under the National Recovery and 
Resilience Plan (NRRP), Mission 4, Component 2, Investment 1.1, funded by the European Union -- 
NextGenerationEU, CUP I53D23000690006.  \\

\end{document}